\documentclass[preprint,eqsecnum,aps,showpacs]{revtex4}
 \usepackage{amssymb} \usepackage{graphicx}

\begin{document}

\title{Wigner functions, coherent states, one-dimensional marginal probabilities and
uncertainty structures of Landau levels}
\author{B. Demircio\u glu$^{1}$}
\email{bengu@taek.gov.tr}
\author{A. Ver\c{c}in$^{2}$}
\email{vercin@science.ankara.edu.tr}
\address{$^{1}$ Sarayk\"{o}y Nuclear Research and Training Center\\
06983, Kazan, Ankara, Turkey\\
$^{2}$ Department of Physics, Ankara University, Faculty of Sciences,\\
06100, Tando\u gan-Ankara, Turkey\\}

\date{\today}

\begin{abstract}
Following an approach based on generating function method phase
space characteristics of Landau system are studied in the autonomous
framework of deformation quantization. Coherent state property of
generating functions is established and marginal probability
densities along canonical coordinate lines are derived. Well defined
analogs of inner product, Cauchy-Bunyakowsy-Schwarz inequality and
state functional have been defined in phase space and they have been
used in analyzing the uncertainty structures. The general form of
the uncertainty relation for two real-valued functions is derived
and uncertainty products are computed in states described by Wigner
functions. Minimum uncertainty state property of the standard
coherent states is presented and uncertainty structures in the case
of phase space generalized coherent states are analyzed.
\end{abstract}
 \maketitle

\section{Introduction and Summary}

 Wigner function is the quantum mechanical analogue of probability
 distribution in a classical phase space and it is the central concept
 of Weyl-Wigner-Groenewold-Moyal quantization \cite{Cahill}. In a
 broader sense this is known also as deformation
 quantization which as an alternative autonomous formulation of
 quantum mechanics has been successfully used in various fields of
 physics and mathematics \cite{Bayen,Zachos1,Zachos2}. Wigner
 function and associated marginal probability (MP) distributions
 have also become important source of information in quantum optics,
 atomic optics and signal processing. In these rapidly developing fields
 experiments aiming to probe the fundamental structure
 and predictions of quantum mechanics by observing
 non-classical behaviors of Wigner functions that are reconstructed
 from measured MP densities along various directions have also been
 designed \cite{Leibfried,Nature,Kim,Bracken}.

 In this paper we study the phase-space properties of Landau
 system \cite{Manko,Vercin2}, that is the motion of a charged particle under the
 influence of a vertical uniform magnetic field, in the
 autonomous framework of deformation quantization without
 using wavefunctions and operators of the conventional quantization
 method. Following a constructive approach based on the generating
 function we will show that basic quantum mechanical
 phase-space characteristics of the problem can be
 completely analyzed. The generating function in deformation
 quantization was firstly introduced in \cite{Zachos3} and
 then used in \cite{Vercin2} in generating Wigner functions and
 two-dimensional (2D) MP densities and in investigating their properties.

Three main contributions of the present study can be summarized as
follows. We first analyze the coherent state property of generating
function and establish main characteristic properties of coherent
states in phase space. Secondly we compute all MP densities along
phase space coordinate lines by introducing new generating
functions. Remarkable properties of the generating functions and
$1D$ MP densities and the fact that they provide important integral
equalities between the classical orthogonal polynomials hardly
obtainable by other means are emphasized. The concentration is then
on analyzing the uncertainty structures in the phase space.
 It is shown that all the mathematical structures and tools such
 as state functional, inner product and related inequalities
 responsible from the uncertainty relations in the Hilbert
 space formulation and forming the basis of many quantum
 mechanical facts, have well defined analogs in deformation quantization.
 To the best of our knowledge, the first
 studies addressing the uncertainty structure in the same context
 are \cite{Zachos4,Przanowski} (see also the recent study \cite{Gerstenhaber}).

 Our approach and results can
 directly be adapted to two mode and extended to multi mode systems of
 quantum optics and they may provide an autonomous phase space
 perspective to our current understanding of quantum Hall
 effects \cite{Aoki} and related issues. Our analysis of
 the uncertainty structures may also have some relevance in the context
 of quantum information processing where a class of inequalities for
 detecting entanglement and for a better understanding of correlations
 has been given in terms of uncertainty relations \cite{Vogel,Hillery}.

 Two main composition rules that will be
 used throughout this study are the $\star$-product and Moyal
 bracket $\{f,g\}_{M}=f\star g-g\star f$ where $f,g$ are arbitrary
 phase-space functions. The $\star $-product is bilinear and
 associative which imply that Moyal bracket is bilinear,
 antisymmetric and obey the Jacobi identity and Leibnitz rule.
 All quantum effects are encoded in these $\hbar$ (the Planck constant)
 dependent composition rules and they obey
\begin{eqnarray}
\lim_{\hbar \rightarrow 0}f\star g=fg\;,\quad \lim_{\hbar \rightarrow 0}\frac{1}{i\hbar
}\{f,g\}_{M} = \{f,g\}\;,\nonumber
\end{eqnarray}
where $fg$ denotes the pointwise product $(fg)(x)=f(x)g(x)$ of $f,g$ and $\{\;,\;\}$
stands for the usual Poisson bracket of the classical mechanics. The above limit
relations constitute the principle of quantum-classical correspondence in the most
concise form at two algebraic levels of observables (real valued phase space functions)
of deformation quantization.

 The phase space in our case is $\mathbb{R}^{4}$
 equipped with canonical coordinates $ {\bf q}=(q_{1},q_{2})$ and
 conjugate momenta ${\bf p}=(p_{1},p_{2})$. The Hamiltonian function
 describing the motion of a spinless particle of charge $q>0$, mass
 $m$ moving on the $q_{1}q_{2}$-plane, reads as
\begin{eqnarray}
H=\frac{1}{2m}({\bf p}-\frac{q}{c}{\bf A})^{2}=\frac{1}{2}%
m(v_{1}^{2}+v_{2}^{2})\;,
\end{eqnarray}
in the Gaussian units. ${\bf A}\equiv {\bf A}({\bf q})$ is the
vector potential of the
 magnetic field $B=\partial _{q_{1}}A_{2}-\partial _{q_{2}}A_{1}$
 which is perpendicular to the plane of motion, $c$ denotes the speed
 of light and $v_{k}$ are the velocity
 components. With the abbreviation
 $\partial_{x_{k}}\equiv\partial /\partial x_{k}$ and
 convention that $\stackrel{\leftarrow }{\partial }$ and
 $\stackrel{\rightarrow }{\partial }$ act, respectively, on the left
 and on the right, the $\star $-product for the present system is
\begin{eqnarray}
 \star =\exp [\frac{1}{2}i\hbar
\sum_{k=1}^{2}(\stackrel{\leftarrow }{
\partial }_{q_{k}}\stackrel{\rightarrow }{\partial }_{p_{k}}-\stackrel{
\leftarrow }{\partial }_{p_{k}}\stackrel{\rightarrow }{\partial
}_{q_{k}})]\;.\label{one}
\end{eqnarray}
Note that for the $k$th star power $(x_{\star})^{k}=x\star\dots
\star x$ (k times) of $x$ we have $(x_{\star})^{k}=x^{k}$ when $x$
is any linear combination of the phase space coordinates.

 Some details of the phase space description of Landau system
 are briefly outlined in the next section where we show how the
 generating function is determined and used in generating Wigner
 functions. For additional details of this section we refer to
\cite{Vercin2,Zachos3}. Coherent state property of the generating
function is specified and
 analyzed in section 3. In section 4 we compute generating
 functions for MP distributions along coordinate lines, derive all
 $1D$ MP densities and then we investigate
their properties.  Construction of several types of integral
 equalities involving the classical orthogonal polynomials are
 also discussed there. The phase space analysis of uncertainty
 structures is taken up in section 5 where the uncertainty
 products in the case of Wigner functions and of standard and
 generalized coherent states are  derived.

\section{Phase-space Description and Generating Function}

For the phase-space quantization of the problem we shall use, when
$B$ is constant, $H$ and
\begin{eqnarray}
J=\frac{1}{2m\omega
}[X_{1}^{2}+X_{2}^{2}-m^{2}(v_{1}^{2}+v_{2}^{2})]\;,\nonumber
\end{eqnarray}
as a set of Moyal-commuting functions. Here $\omega =qB/mc$ is the
cyclotron frequency and $X_{1}=m(v_{2}+\omega q_{1}),\;X_{2}=
-m(v_{1}-\omega q_{2})$ are constants of motion:
$\{H,X_{k}\}_{M}=0$. $X_{k}$'s are proportional to the coordinates
of cyclotron center and they satisfy the gauge-independent relation
$\{X_{1},X_{2}\}_{M}=-im\hbar\omega$. $J$ corresponds to canonical
angular momentum $q_{1}p_{2}-q_{2}p_{1}$ in the henceforth assumed
symmetric gauge $ {\bf A}=B(-q_{2},\;q_{1},\;0)/2$. Representing the
complex conjugation by overbar and the magnetic length $(2\hbar
/m\omega)^{1/2}$ by $\gamma$, two mutually commuting pairs of
dimensionless creation $\bar{a},\bar{b}$ and annihilation functions
\begin{eqnarray}
a &=&\frac{1}{\gamma \omega }(v_{1}+iv_{2})=\frac{1}{m\gamma \omega
}(p_{1}+ip_{2})-\frac{i}{2\gamma}(q_{1}+iq_{2})\;, \\
b &=&\frac{1}{m\gamma \omega }(X_{2}+iX_{1})=-\frac{1}{m\gamma \omega }(p_{1}-ip_{2})+
\frac{i}{2\gamma}(q_{1}-iq_{2})\;,
\end{eqnarray}
which satisfy $\{a,\bar{a}\}_{M} =1=\{b,\bar{b}\}_{M}$ can be
defined. These enable us to rewrite  ~(\ref{one}) as
\begin{eqnarray}
\star =exp[\frac{1}{2}(\stackrel{\leftarrow }{\partial }_{a}\stackrel{%
\rightarrow }{\partial }_{\bar{a}}+\stackrel{\leftarrow }{\partial }_{b}%
\stackrel{\rightarrow }{\partial }_{\bar{b}}-\stackrel{\leftarrow
}{\partial
}_{\bar{a}}\stackrel{\rightarrow }{\partial }_{a}-\stackrel{\leftarrow }{%
\partial }_{\bar{b}}\stackrel{\rightarrow }{\partial }_{b})]\;.
\end{eqnarray}

Higher level Wigner functions can be generated by successive
application of the creation functions (and of the annihilation
functions from the right) to the ground state Wigner function
$W_{0}$. An efficient way of achieving this goal is to introduce a
generating function which can be inferred from $W_{0}$. Defining
equations of $W_{0}$ are $a\star W_{0} = 0 = b\star W_{0}$ whose
real solution, normalized with the volume
 element $dV =dq_{1}dq_{2}dp_{1}dp_{2}$, is
\begin{eqnarray}
W_{0}=4e^{-2(a\bar{a}+b\bar{b})}\;,\quad
\int_{\mathbb{R}^{4}}W_{0}dV=h^{2}\;.\nonumber
\end{eqnarray}
We now introduce, in terms of complex parameters
$\alpha_{k},\beta_{k}$ the phase-space functions
\begin{eqnarray}
G_{1} =G_{1}(a,\bar{a};\alpha_{1},\beta_{1})=e^{\alpha_{1} \bar{a}}\star
e^{-2a\bar{a}}\star e^{\beta_{1}a},
\end{eqnarray}
and $G_{2}=G_{2}(b,\bar{b};\alpha_{2},\beta_{2})$. Since $G_{1}=
\exp[-\alpha _{1}\beta _{1}+2(\alpha _{1}\bar{a} +\beta
_{1}a-a\bar{a})]$, we obtain
\begin{eqnarray}
G_{1}= e^{-2a\bar{a}}\sum_{k,n=0}^{\infty }\frac{%
\alpha _{1}^{k}}{k!}(2\bar{a})^{k-n}(-\beta
_{1})^{n}L_{n}^{k-n}(4a\bar{a})\;,\nonumber
\end{eqnarray}
and a similar relation for $G_{2}$ by recalling the definition of
the generalized Laguerre polynomials:
$(1+y)^{k}e^{-xy}=\sum_{n=0}L_{n}^{k-n}(x)y^{n}$ \cite{Magnus}. As
the generating function we take $G=G_{1}G_{2}$.

Finally in this section let us consider the phase-space functions
\begin{eqnarray}
w_{n_{1}n_{2}}&=& N_{n}\partial^{n_{1}}_{\alpha_{1}}
\partial^{n_{2}}_{\beta_{1}}G_{1}|_{\alpha_{1}=0=\beta_{1}}
=N_{n} \bar{a}^{n_{1}}\star e^{-2a\bar{a}}\star a^{n_{2}}\;\nonumber\\
w_{\ell_{1}\ell_{2}}&=&N_{\ell}
\partial^{\ell_{1}}_{\alpha_{2}} \partial^{\ell_{2}}_{\beta_{2}} G_{2}|_{\alpha_{2}
=0=\beta_{2}}=N_{\ell} \bar{b}^{\ell_{1}}\star e^{-2b\bar{b}}\star
b^{\ell_{2}}\;\nonumber
\end{eqnarray}
where $n_{k},l_{k}$ are positive integers and $N_{n}=
(n_{1}!n_{2}!)^{-1/2},\;N_{\ell}=(\ell_{1}!\ell_{2}!)^{-1/2}$ . Then
all Wigner functions can be constructed from
$W_{n_{1}n_{2}\ell_{1}\ell_{2}}= 4w_{n_{1}n_{2}}
w_{\ell_{1}\ell_{2}}$ whose special cases for $n_{1}=n_{2}=n$ and
$\ell_{1}=\ell_{2}=\ell$ correspond to the diagonal (or pure state)
Wigner functions
\begin{eqnarray}
W_{n\ell}=\frac{1}{n ! \ell !} \bar{a}^{n}\star \bar{b}^{\ell}\star
W_{0} \star a^{n}\star
b^{\ell}=(-1)^{n+\ell}L_{n}(4a\bar{a})L_{\ell}(4b\bar{b})W_{0}\;,
\end{eqnarray}
with $W_{0}\equiv W_{00}$. It is now easy to check the
$\star$-ladder structures; $a\star W_{n\ell}=W_{n-1, \ell}\star a$
and $\bar{a}\star W_{n\ell} = W_{n+1,\ell}\star \bar{a}$ which
implies $n_{a}\star W_{n\ell} = W_{n\ell}\star n_{a} = nW_{n\ell}$
for real number function $n_{a}=\bar{a}\star a$. Similar relations
hold for $b,\;\bar{b}$ and $n_{b}=\bar{b}\star b$. These justify the
fact that $\{W_{n\ell}; \;n,\ell=0,1,2,...\}$ is the set of
simultaneous $\star$-eigenfunctions of the so-called two-sided
$\star$-eigenvalue equations $H\star W_{n\ell}=W_{n\ell}\star
H=E_{n}W_{n\ell}$ for $H = \hbar \omega (2n_{a}+1)/2$ and the
similar one for $J=\hbar (n_{b}-n_{a})$ with eigenvalues
\begin{eqnarray}
E_{n}=\hbar \omega (n+\frac{1}{2})\;,\quad J_{n\ell} =\hbar
(\ell-n)\;.
\end{eqnarray}
$E_{n}$ are the well-known infinitely degenerate (for they are
independent from $\ell$) Landau levels.

\section{Phase-Space Coherent States}

In this section we shall establish another essential property of
generating function
\begin{equation}
G=G_{1}G_{2}=e^{-\alpha \cdot \beta} e^{2(\alpha _{1}\bar{a%
}+\beta _{1}a+\alpha _{2}\bar{b}+ \beta
_{2}b)}e^{-2(a\bar{a}+b\bar{b})}\;,
\end{equation}
where $\alpha\cdot\beta=\alpha _{1}\beta_{1}+\alpha _{2}\beta _{2}$
. This is the fact that $G$ with its complex parameters represents
the standard phase-space coherent states of Landau system. The
standard coherent states can be defined by three interrelated ways
\cite{Perelomov,Gilmore} (and for a recent review \cite{Vourdas})
which, by adopting them for the deformation quantization, can be
stated as follows. (i) They are the simultaneous one sided
star-eigenfunctions of annihilation functions. (ii) They can be
generated by application of the phase space displacement function
(introduced below) to the ground state Wigner function. (iii) They
are phase space functions with the minimum uncertainty relationship.

Modern group theoretical descriptions of coherent states were given
by Perelomov who also generalized the standard coherent states first
purposed by Schr\"{o}dinger and then revived by Glauber with
important applications in quantum optics. Phase space analogue of the
Perelomov generalized coherent states and their uncertainty structures
will be presented, together of point (iii) mentioned above, in the
last section there we first discuss how to analyze the uncertainty
structures in a classical phase space.

The most direct way of exhibiting the coherent state property of $G$
is to show that it is a left $\star$-eigenfunction of both $a$ and
$b$. This can be easily seen, by Eq. (2.4), from
\begin{eqnarray}
a\star G=(a+\frac{1}{2}\partial _{\bar{a}})G =\alpha _{1}G\;,\quad
 G\star \bar{a}
=(\bar{a}+\frac{1}{2}\partial _{a})G =\beta _{1}G\;.
\end{eqnarray}
Similar relations hold for $b$ and $\bar{b}$. Thus $G$ behaves as a
left/right coherent state with (one-sided) $\star$-eigenvalues
$\alpha_{k}$ and $\beta_{k}$. For real $G$ we take
$\beta=\bar{\alpha}$. In such a case $G$ corresponds to the
Glauber-Perelomov standard coherent state and reads from (2.4) and
(3.1) as
\begin{eqnarray}
G^{\prime}&=&\frac{1}{4} e^{\alpha _{1}\bar{a}}\star e^{\alpha _{2}\bar{b}}\star
W_{0}\star e^{\bar{\alpha}_{1}a}\star e^{\bar{\alpha}_{2}b}\nonumber\\
&=&\frac{1}{4}e^{-|\alpha_{1}|^{2}-|\alpha_{2}|^{2}} e^{2(\alpha
_{1}\bar{a}+\bar{\alpha}_{1}a+\alpha _{2}\bar{b}+\bar{\alpha} _{2}b)}W_{0}.
\end{eqnarray}
Note that among all Wigner functions only $W_{0}$ is a (normalized)
coherent state.

We will now show that $G^{\prime}$ can be defined by application of a phase space
displacement function to the ground state Wigner function $W_{0}$. For this purpose we
first observe that
\begin{eqnarray}
e^{\eta a}\star e^{-2a\bar{a}}&=&e^{\eta a}e^{\frac{1}{2} \stackrel{\leftarrow
}{\partial }_{a}\stackrel{\rightarrow }{\partial }
_{\bar{a}}}e^{-2a\bar{a}}\;,\nonumber\\
&=&\sum _{n=0}^{\infty}\frac{1}{n!}(\frac{1}{2})^{n}({\partial }_{a}^{n}e^{\eta
a})(\partial _{\bar{a}}^{n}e^{-2a\bar{a}})=e^{-2a\bar{a}}\;.\nonumber
\end{eqnarray}
Similar calculations show that $e^{-2a\bar{a}}$ is also a right $\star$-eigenfunction
of $e^{\kappa\bar{a}}$, hence
\begin{eqnarray}
e^{\eta a}\star e^{-2a\bar{a}}\star e^{\kappa \bar{a}}=e^{-2a\bar{a}}\;.
\end{eqnarray}
This may be inferred from the definition of $W_{0}$. We now introduce the
complex-valued displacement function
$D_{1}=D_{1}(a,\bar{a},\alpha_{1},\bar{\alpha}_{1})$
\begin{eqnarray}
D_{1}&=&e^{\alpha_{1}\bar{a}-\bar{\alpha}_{1}a}\;\\
&=& e^{-|\alpha_{1}|^{2}/2} e^{\alpha_{1}\bar{a}}\star
e^{-\bar{\alpha}_{1}a}=e^{|\alpha_{1}|^{2}/2} e^{-\bar{\alpha}_{1}a}\star e^{\alpha_{1}
\bar{a}}\;.\nonumber
\end{eqnarray}
For the factorizations in the second line we made use of
\begin{eqnarray}
e^{\eta a}\star e^{\kappa\bar{a}}=e^{\eta\kappa/2} e^{\eta a+\kappa
\bar{a}}=e^{\eta\kappa}e^{\kappa \bar{a}}\star e^{\eta a}\;.\nonumber
\end{eqnarray}

$\bar{D}_{1}$ being the complex conjugate of $D_{1}$, from Eqs. (3.4) and (3.5) we
obtain
\begin{eqnarray}
D_{1}\star e^{-2a\bar{a}}\star \bar{D}_{1}= e^{-|\alpha_{1}|^{2}}
e^{\alpha_{1}\bar{a}}\star e^{-2a\bar{a}}\star e^{\bar{\alpha}_{1} a}\;.\nonumber
\end{eqnarray}
By defining $D_{2}=D_{2}(b,\bar{b},\alpha_{2},\bar{\alpha}_{2})$ and
taking the displacement function as
\begin{eqnarray}
D_{\alpha_{1}\alpha_{2}}=D_{1}D_{2}\;,
\end{eqnarray}
the desired result is achieved as follow
\begin{eqnarray}
e^{-|\alpha_{1}|^{2}-|\alpha_{2}|^{2}}G^{\prime}=\frac{1}{4}D_{\alpha_{1}\alpha_{2}}\star
W_{0}\star \bar{D}_{\alpha_{1}\alpha_{2}}\;.
\end{eqnarray}
That is, up to a positive constant, $G^{\prime}$ can be defined
by left-right action of the phase-space displacement function
to $W_{0}$. Like $W_{0},\;G^{\prime}$ is positive-valued at each
point of the phase space and for all values of the complex
parameters $\alpha_{k}$. Note also that $D_{k}(x)$, where $k=1,2$
and $x=(a,\bar{a})$ or $x=(b,\bar{b})$, are $\star$-unitary
phase space functions in the sense that $\bar{D}_{k}(x)=D_{k}(-x)$ and
\begin{eqnarray}
D_{k}(x)\star \bar{D}_{k}(x)=1=\bar{D}_{k}(x)\star D_{k}(x)\;,\nonumber
\end{eqnarray}
which imply
\begin{eqnarray}
D_{\alpha_{1}\alpha_{2}}\star
\bar{D}_{\alpha_{1}\alpha_{2}}=1=\bar{D}_{\alpha_{1}\alpha_{2}}\star
D_{\alpha_{1}\alpha_{2}}\;.
\end{eqnarray}

To emphasize the displacement property of $D_{\alpha_{1}\alpha_{2}}$, we first observe
that
\begin{eqnarray}
\;\{ a,D_{\alpha_{1}\alpha_{2}} \}_{M}&=&\alpha_{1}
D_{\alpha_{1}\alpha_{2}}\;,\nonumber\\
 \{\bar{a},D_{\alpha_{1}\alpha_{2}}\}_{M}&=&\bar{\alpha}_{1}
 D_{\alpha_{1}\alpha_{2}}\;,\nonumber\\
\{b,D_{\alpha_{1}\alpha_{2}}\}_{M}&=&\alpha_{2}
D_{\alpha_{1}\alpha_{2}}\;,\nonumber\\
\{\bar{b},D_{\alpha_{1}\alpha_{2}}\}_{M}&=&\bar{\alpha}_{2}
D_{\alpha_{1}\alpha_{2}}\;,\nonumber
\end{eqnarray}
and then we obtain (complex conjugation gives similar relations for $\bar{a}$ and
$\bar{b}$)
\begin{eqnarray}
\bar{D}_{\alpha_{1}\alpha_{2}}\star a\star D_{\alpha_{1}\alpha_{2}}&=&
a+\alpha_{1}\;,\nonumber\\
\bar{D}_{\alpha_{1}\alpha_{2}}\star b\star D_{\alpha_{1}\alpha_{2}}&=&
b+\alpha_{2}\;.\nonumber
\end{eqnarray}
For generalization suppose that $f$ is a phase space function that can be expanded in a
star power series of creation and annihilation functions such as
\begin{eqnarray}
f=f(a,\bar{a},b,\bar{b})=\sum_{jj^{\prime}kk^{\prime}}c_{jj^{\prime}kk^{\prime}}a^{j}\star\bar{a}^{j^{\prime}}\star
b^{k}\star \bar{b}^{k^{\prime}}\;,\nonumber
\end{eqnarray}
where $c_{jj^{\prime}kk^{\prime}}$'s are some constants and all the indices  take
positive integer values. Then the displaced function $f^{\prime}$ of $f$ defined by
\begin{eqnarray}
f^{\prime}=\bar{D}_{\alpha_{1}\alpha_{2}}\star f \star D_{\alpha_{1}\alpha_{2}}\;,
\end{eqnarray}
is of the form
\begin{eqnarray}
f^{\prime}(a,b,\bar{a},\bar{b})=f(a+\alpha_{1},b+\alpha_{2},\bar{a}+\bar{\alpha_{1}},
\bar{b}+\bar{\alpha_{2}})\;.
\end{eqnarray}

 Up to now two essential properties of $G$ have been established. The
 first employed in the previous section was that all (diagonal
 and off-diagonal) Wigner functions can be generated from $G$. The
 second presented above emphasize the phase space coherent state
 property of $G$. As has been shown to each point $(\alpha_{1},
 \alpha_{2})$ of a $2D$ complex space $G^{\prime }$ assigns a
 real-valued, phase space coherent state function of the Landau
 system. In the next section another basic property
 of $G$ is established. This is the fact that integrated forms of $G$ serve as
 generating functions for MP densities.

\section{1D Marginal
Probability Densities}

For the third property of $G$ stated above we take multiple integral of $G$ on various
phase-space regions. As an example let us consider the function of $q_{1},q_{2}$
defined by
\begin{equation}
M_{\alpha \beta}(q_{1},q_{2}) =
\int_{\mathbb{R}^{2}}Gdp_{1}dp_{2}=\frac{\pi\hbar^{2}}{\gamma^{2}}
e^{- \alpha_{1}\alpha_{2} -\beta_{1}\beta_{2}}e^{i(\alpha
_{1}\bar{Z}-\alpha _{2}Z)-i(\beta
_{1}Z-\beta_{2}\bar{Z})}e^{-Z\bar{Z}}\;,
\end{equation}
where $Z=(q_{1}+iq_{2})/\gamma$.  We derive in terms of it
\begin{equation}
P_{n\ell}(q_{1}, q_{2})=\frac{4}{n!\ell!} (\partial_{\alpha_{1}}
\partial_{\beta_{1}})^{n} (\partial_{\alpha_{2}}\partial_{\beta_{2}})^{\ell}
M_{\alpha \beta}(q_{1},q_{2})
|_{\bbox{\alpha}=\bbox{0}=\bbox{\beta}}.
\end{equation}
Combining these two relations and comparing the result with the
definition of Wigner functions given by (2.7) we obtain
$P_{n\ell}(q_{1}, q_{2})=\int_{\mathbb{R}^{2}}W_{nl}dp_{1}dp_{2}$.
That is, $P_{n\ell}(q_{1}, q_{2})$ is indeed the MP density in the
$q_{1}q_{2}$-plane and $M_{\alpha \beta}$ plays the role of
generating function for it.

\subsection{Generating Functions For $1D$ Marginal Probability Densities}

Integrating $M_{\alpha \beta}(x_{i},x_{j})$ on one of its coordinate
gives the generating function depending on the remaining coordinate.
There are three possibilities in each case that can be utilized for
checking the calculations. As an example, the generating function
for the MP densities in $q_{1}$-direction can be computed as
\begin{eqnarray}
Q_{\alpha \beta}(q_{1})=\int_{-\infty}^{\infty}M_{\alpha
\beta}(q_{1},q_{2})
dq_{2}=\int_{\mathbb{R}^{3}}Gdp_{1}dp_{2}dq_{2}\;,\nonumber
\end{eqnarray}
This can equivalently be obtained from the integral of $M_{\alpha \beta}(q_{1},p_{1})$
or of $M_{\alpha \beta}(q_{1},p_{2})$ over $p_{1}$ and $p_{2}$, respectively. All
generating functions have been computed in this way and they are found to be
\begin{mathletters}
\begin{eqnarray}
Q_{\alpha\beta}(q_{1})&=&N_{q}
e^{\bbox{\alpha\cdot\beta}}\exp{\{-[\frac{1}{\gamma}q_{1}-\frac{i}{2}(\alpha
_{1}-\alpha _{2}-\beta _{1}+\beta _{2})]^{2}}\}\;,\nonumber\\
Q_{\alpha \beta}(p_{1}) &=& N_{p} e^{\bbox{\alpha\cdot\beta}}
\exp{\{-[\frac{\gamma}{\hbar} p_{1}-\frac{1}{2}(\alpha _{1}-\alpha
_{2}+\beta _{1}-\beta _{2})]^{2}}\}\;,\\
Q_{\alpha \beta}(q_{2}) &=& N_{q} e^{\bbox{\alpha\cdot\beta}}
\exp{\{-[{\frac{1}{\gamma}q_{2}}-{\frac{1}{2}}(\alpha _{1}+\alpha
_{2}+\beta _{1}+\beta _{2})]^{2}}\}\;,\nonumber\\
Q_{\alpha \beta}(p_{2})&=& N_{p} e^{\bbox{\alpha\cdot\beta}} \exp{\{-[{\frac{
\gamma}{\hbar}p_{2}}+{\frac{i}{2}}(\alpha _{1}+\alpha _{2}-\beta _{1}-\beta
_{2})]^{2}}\}\;,\nonumber
\end{eqnarray}
\end{mathletters}
where
\begin{eqnarray}
N_{q}= \frac{\pi^{3/2} \hbar^{2}}{\gamma},\quad N_{p}= \pi^{3/2}\hbar\gamma\;.
\end{eqnarray}

In the next subsection, $1D$ MP densities in $x_{i}$-direction will
be found from
\begin{eqnarray}
P_{n\ell}(x_{i})=\frac{4}{n!\ell!} (\partial_{\alpha_{1}}
\partial_{\beta_{1}})^{n} (\partial_{\alpha_{2}}\partial_{\beta_{2}})^{\ell}
Q_{\alpha \beta}(x_{i}) |_{\bbox{\alpha}=\bbox{0}=\bbox{\beta}}.
\end{eqnarray}
Now by comparing this equation with
\begin{equation}
P_{n\ell}(q_{1})=\int_{\mathbb{R}^{3}}W_{nl}dp_{1}dp_{2}dq_{2} \;,
\end{equation}
 one of the advantages of the generating function method can easily
 be recognized. Instead of taking multiple integral of special functions
 in deriving MPs from (4.6) and (2.7) one can obtain them more easily
 by taking derivatives of a exponential function given by one of
 (4.3). $2D$ probability densities for the phase-space planes
 were derived \cite{Vercin2} in a similar way.

 \subsection{Derivation of $1D$ Marginal Probability Densities}

By defining $u=\exp(\alpha\cdot\beta)$ and $v=\exp(-z^2)$ with,
\begin{eqnarray}
z=\frac{1}{\gamma}q_{1}-\frac{i}{2}(\alpha _{1}-\alpha _{2}-\beta _{1}+\beta _{2})\;,
\end{eqnarray}
we can write $Q_{\alpha\beta}(q_{1})=N_{q}uv$. We then obtain $P_{n\ell}(q_{1})$ in two
steps. Firstly we compute
\begin{eqnarray}
I_{1}=[(\partial_{\beta_{1}}\partial_{\alpha_{1}})^{n} uv]|_{\alpha_{1}=0=\beta_{1}},
\end{eqnarray}
and then the result will be read, in view of (4.5), from
\begin{eqnarray}
P_{n\ell}(q_{1})=\frac{4}{n!\ell!}
N_{q}(\partial_{\alpha_{2}}\partial_{\beta_{2}})^{\ell}
I_{1}|_{\alpha_{2}=0=\beta_{2}}.
\end{eqnarray}

Using the Leibnitz rule
\begin{eqnarray}
\partial_{x}^{n}
(uv)=\sum_{j=0}^{n}\left( {\begin{array}{c}
n \\
j
\end{array}}
\right)\partial_{x}^{j}u \partial_{x}^{n-j}v\;,\nonumber
\end{eqnarray}
and $\partial_{x}^{j} x^{n}=n!x^{n-j}/(n-j)!$, for $j\leq n$, we get by direct
computations
\begin{eqnarray}
(\partial_{\beta_{1}}\partial_{\alpha_{1}})^{n} uv
&=&\partial_{\beta_{1}}^{n}[u\sum_{j=0}^{n}\left( {\begin{array}{c}
n \\
j
\end{array}}
\right)\beta_{1}^{j}(-\frac{i}{2}\partial_{z})^{n-j}v]
\nonumber \\
&=&u \sum_{k=0}^{n}\left( {\begin{array}{c}
n \\
k
\end{array}}
\right)\alpha_{1}^{k}\partial_{\beta_{1}}^{n-k} [\sum_{j=0}^{n}\left( {\begin{array}{c}
n \\
j
\end{array}}
\right)\beta_{1}^{j}(-\frac{i}{2}\partial_{z})^{n-j}v]\;,\nonumber\\
&=&u\sum_{k=0}^{n}\sum_{j=0}^{n}\sum_{s=0}^{n-k\;\prime}
B_{nkjs}\alpha_{1}^{k}\beta_{1}^{j-s}
\partial_{z}^{2n-k-s-j}{v}\;,
\end{eqnarray}
where prime over $s-$summation indicates the restriction $s\leq j$ and
\begin{eqnarray}
B_{nkjs}=\left( {\begin{array}{c}
n \\
k
\end{array}}
\right)\left( {\begin{array}{c}
n \\
j
\end{array}}
\right)\left( {\begin{array}{c}
n-k \\
s
\end{array}}
\right)(-1)^{n-j}(\frac{i}{2})^{2n-k-s-j}\frac{j!}{(j-s)!}. \nonumber
\end{eqnarray}
In the first and last line of (4.10), $\partial_{\alpha_{1}}$ and
$\partial_{\beta_{1}}$ are taken, in view of (4.7), as
$\partial_{\alpha_{1}}=-i\partial_{z}/2$ and
$\partial_{\beta_{1}}=i\partial_{z}/2$ when they are acting on $v$.
Evaluating (4.10) at $\alpha_{1}=0=\beta_{1}$ amounts to taking
$k=0$ and $s=j$. Hence $I_{1}=e^{\alpha_{2}\beta_{2}}g(z_{1})$ where
$z_{1}=\frac{1}{\gamma}q_{1}+\frac{i}{2}(\alpha _{2}-\beta _{2})$
and
\begin{eqnarray}
g(z_{1})=\sum_{j=0}^{n}\left( {\begin{array}{c}
n \\
j
\end{array}}
\right)^{2}(\frac{1}{4})^{n-j}j!\partial_{z_{1}}^{2(n-j)}{e^{-z_{1}^{2}}}\;.
\end{eqnarray}

In the second step we compute, in a similar way
\begin{eqnarray}
I_{2}=(\partial_{\beta_{2}}\partial_{\alpha_{2}})^{\ell}I_{1}=
e^{\alpha_{2}\beta_{2}}\sum_{r=0}^{\ell}\left( {\begin{array}{c}
\ell \\
r
\end{array}}
\right){\alpha_{2}}^{r}{\partial_{\beta_{2}}^{\ell-r}}[\sum_{k=0}^{\ell}\left(
{\begin{array}{c}
\ell \\
k
\end{array}}
\right){\beta_{2}}^{k}(\frac{i}{2}\partial_{z_{1}})^{\ell-k} g(z_{1})].
\end{eqnarray}
For $\alpha_{2}=0$ this transforms to
\begin{eqnarray}
I_{2}|_{\alpha_{2}=0} &=&\partial_{\beta_{2}}^{\ell}[\sum_{k=0}^{\ell}\left(
{\begin{array}{c}
\ell \\
k
\end{array}}
\right)(\frac{i}{2})^{\ell-k}{\beta_{2}}^{k}\partial_{z_{2}}^{\ell-k}
g(z_{2})]\nonumber\\
 &=&\sum_{k=0}^{\ell}\sum_{t=0}^{\ell}\left(
{\begin{array}{c}
\ell \\
k
\end{array}}
\right)\left( {\begin{array}{c}
\ell
\\
\end{array}}
\right)(-1)^{\ell-t}(\frac{i}{2})^{2\ell-k-t}\frac{k!}{(k-t)!}{\beta_{2}}^{k-t}
\partial_{z_{2}}^{2\ell-k-t} g(z_{2})\;,\nonumber
\end{eqnarray}
where $z_{2}=z_{1}|_{\alpha_{2}=0}$. Evaluating for $\beta_{2}=0$ it simplifies to
\begin{eqnarray}
I_{2}|_{\alpha_{2}=0=\beta_{2}}&=&\sum_{k=0}^{\ell}\left( {\begin{array}{c}
\ell \\
k
\end{array}}
\right)^{2}k!(\frac{1}{2}\partial_{y})^{2(\ell-k)}g(y)\nonumber\\
&=&\sum_{j=0}^{n}\sum_{k=0}^{\ell} \left( {\begin{array}{c}
n \\
j
\end{array}}
\right)^{2} \left( {\begin{array}{c}
\ell \\
k
\end{array}}
\right)^{2}j!k!(\frac{1}{2}\partial_{y})^{2(n+\ell-j-k)}e^{-y^{2}}\;,
\end{eqnarray}
 where $y=z_{2}|_{\beta_{2}=0}=q_{1}/ \gamma$. Finally using the Rodriguez
 formula for Hermite polynomials
\begin {eqnarray}
H_{n}(y)=(-1)^{n}e^{y^{2}}\partial_{y}^{n}e^{-y^{2}}\;,
\end{eqnarray}
MP density in $q_{1}$-direction are found from (4.9) to be
\begin {eqnarray}
P_{n\ell}(q_{1})= N_{q}e^{-q_{1}^{2}/
\gamma^{2}}\sum_{j=0}^{n}\sum_{k=0}^{\ell}A_{n\ell
jk}H_{2(n+\ell-j-k)}(\frac{q_{1}}{\gamma}).
\end{eqnarray}
where we have defined
\begin{eqnarray}
A_{n\ell jk}= 4{\frac{j!k!}{n!\ell!}} \left( {\begin{array}{c}
n \\
j
\end{array}}
\right)^{2}\left( {\begin{array}{c}
\ell \\
k
\end{array}}
\right)^{2} (\frac{1}{4})^{n+\ell-j-k}\;.
\end{eqnarray}

The calculations for other coordinates have been performed as well and the results are
given altogether as follows ($i=1,2$)
\begin{eqnarray}
P_{n\ell}(q_{i}) &=& N_{q}e^{-q_{i}^{2}/ \gamma^{2}} \sum_{j=0}^{n} \sum_{k=0}^{\ell}
A_{n\ell
jk}H_{2(n+\ell-j-k)}(\frac{q_{i}}{\gamma})\;,\\
P_{n\ell}(p_{i})&=& N_{p}e^{-\gamma^{2} p_{i}^{2}/ \hbar^{2}} \sum_{j=0}^{n}
\sum_{k=0}^{\ell} A_{n\ell jk}H_{2(n+\ell-j-k)}(\frac{\gamma p_{i}}{\hbar})\;.
\end{eqnarray}
The rest of the section is devoted to investigation of some
properties of these $1D$ MPs.

\subsection{Symmetry and Normalization}

Like Wigner functions $1D$ MP distributions are localized around the
origin and they are even functions of the corresponding coordinates.
Another observation, as is obvious from Eqs. (4.17) and (4.18), is
the symmetry property in quantum numbers $n$ and $\ell$
\begin{eqnarray}
P_{n\ell}(x)=P_{\ell n}(x)\;.
\end{eqnarray}
It follows from Eqs. (2.6) that the quantum number $n$ determines
the energy levels and hence the radius of cyclotron while $\ell$
determines, together with $n$, the angular momentum states with
$(\ell-n)$. On the other hand, from the content of section II it is
not hard to verify that the distance of cyclotron center from the
origin is specified by $\ell$ itself. That is, Eq. (4.19) simply
says that $1D$ position and momentum probability distributions are
symmetric with respect to these two distances and one can not
distinguish them from a given $P_{n\ell}(x)$.

As a consistency check we can easily show that $P_{n\ell}(x_{i})$ are
all normalized as follow
\begin{eqnarray}
\int_{-\infty}^{\infty}P_{n\ell}(q_{1})dq_{1} &=& N_{q}\sum_{j=0}^{n} \sum_{k=0}^{\ell}
A_{n\ell jk} \int_{-\infty}^{\infty}e^{-q_{1}^{2}/ \gamma^{2}}
H_{2(n+\ell-j-k)}(\frac{q_{1}}{\gamma})dq_{1}\nonumber\\
&=& 4N_{q}\gamma\pi^{1/2}=h^{2}\;.
\end{eqnarray}
Noting that $H_{0}(x)=1$, this follows from the orthogonality relation
\begin{eqnarray}
\int_{-\infty}^{\infty}e^{-x^{2}}H_{m}(x)H_{n}(x)dx=\pi^{1/2}2^{n}n!\delta_{nm}\;.
\end{eqnarray}
More explicitly, for $y=q_{1}/\gamma$ and for non-zero positive integer $m$,
we get
\begin{eqnarray}
\int_{-\infty}^{\infty}e^{-y^{2}}H_{2m}(y)dq_{1}=2\gamma\int_{0}^{\infty}\frac
{\partial^{2m}}{\partial y^{2m}}e^{-y^{2}}dy=2\gamma H_{2m-1}(0)=0\;,\nonumber
\end{eqnarray}
for odd-parity Hermite polynomials vanishes at the origin. Hence
only $j=n,\;k=\ell$ term in (4.20) contributes to the normalization.
Therefore  $A_{n\;\ell\; n\; \ell}=4$ and the well-known equality
$\int_{-\infty}^{\infty}e^{-y^{2}}dq_{1}=\gamma\pi^{1/2}$ prove the
result. Note that (4.20) proves the normalization of all Wigner
functions and this is a direct result of normalization of $W_{0}$.

\subsection{Integral Equalities}

 Having a complete list of Wigner functions and of associated $1D$ MPs at
 hand a generic type of integral equality in the context of the theory
 of orthogonal polynomials can be written from Eq. (4.6). Similarly
 $P_{n\ell}(q_{1}, q_{2})=\int_{\mathbb{R}^{2}}W_{nl}dp_{1}dp_{2}$
 gives different type of equality by using $2D$ MPs
 $P_{nl}(x_{i}, x_{j})$ found in \cite{Vercin2}. In fact by using
\begin{eqnarray}
P_{nl}(q_{1}, q_{2})&=&N_{nl} (\frac{\hbar}{\gamma})^{2}\rho^{2(n-l)}e^{-%
\rho^{2}} [L^{n-l}_{l}(\rho^{2})]^{2}\;,\nonumber\\
P_{nl}(q_{1}, p_{2})&=& N^{\prime}_{nl} \hbar
e^{-\frac{1}{2}(\tau^{2}_{+}+\tau^{2}_{-}) } H^{2}_{n}( \frac{\tau_{-}}{\sqrt{2}})
H^{2}_{l}(\frac{\tau_{+}}{\sqrt{2}}) \;,
\end{eqnarray}
where $\rho^{2}=Z\bar{Z},\;N_{nl}=4\pi l!/n!\;,N_{nl}^{\prime }=4\pi /n!l!2^{n+l}$ and
\begin{eqnarray}
\zeta^{2}=\frac{\gamma^{2} (p_{1}^{2}+p_{2}^{2}) }{ 4\hbar^{2}}\;,
\quad\tau_{\pm}
&=& \frac{q_{1}}{\gamma}\pm \frac{\gamma p_{2}}{\hbar}\;.\nonumber
\end{eqnarray}
two additional type of equalities can easily be obtained from
\begin{eqnarray}
P_{n\ell}(x_{i})=\int_{-\infty}^{\infty}P_{nl}(x_{i},\;x_{j})dx_{j}\;.
\end{eqnarray}
Indeed, when (4.17) and (4.22) are substituted in (4.23) we get,
after canceling the $\exp(-q_{1}^{2}/\gamma^{2})$ from both sides,
the following generic type of integral equalities
\begin{eqnarray}
\sum_{j=0}^{n} \sum_{k=0}^{\ell} A_{n\ell
jk}H_{2(n+\ell-j-k)}(\frac{q_{1}}{\gamma})&=&\frac{N_{nl}}{N_{q}}
(\frac{\hbar}{\gamma})^{2}\int_{-\infty}^{\infty}\rho^{2(n-l)}e^{-q_{2}^{2}/\gamma^{2}} [L^{n-l}_{l}(\rho^{2})]^{2}dq_{2}\;,\nonumber\\
&=&\frac{N^{\prime}_{nl}}{N_{q}} \hbar\int_{-\infty}^{\infty}
e^{-\gamma^{2}p_{2}^{2}/\hbar^{2} } H^{2}_{n}( \frac{\tau_{-}}{\sqrt{2}})
H^{2}_{l}(\frac{\tau_{+}}{\sqrt{2}})dp_{2} \;,
\end{eqnarray}
which can hardly be obtained by other means.

Despite the fact that even parity polynomials $H_{2m}$ take positive
as well as negative values $P_{n\ell}(x_{i})$ are always positive
valued. This follows from the fact that, like that given by (4.22),
all $P_{nl}(x_{i}, x_{j})$'s (and hence their integral on of the
coordinate lines) are positive on the corresponding phase space
planes. Finally in this section we give explicit forms of
$P_{n\ell}(q_{1})$ for some low lying states
\begin{eqnarray}
P_{00}(q_{1})&=&4N_{q}e^{-y^{2}}\;,\nonumber\\
P_{10}(q_{1})&=&2N_{q}e^{-y^{2}}(2y^{2}+1)\;,\nonumber\\
P_{11}(q_{1})&=&N_{q}e^{-y^{2}}(4y^{4}-4y^{2}+3)\;,\nonumber\\
P_{20}(q_{1})&=&\frac{1}{2}N_{q}e^{-y^{2}}(4y^{4}+4y^{2}+3)\;,\nonumber\\
P_{21}(q_{1})&=&\frac{1}{4}N_{q}e^{-y^{2}}(8y^{6}-20y^{4}+18 y^{2}+7)\;. \nonumber
\end{eqnarray}

\section{Uncertainty Structures in the Phase Space}

In this section we show that all the uncertainty structures of the quantum mechanics
can be realized in a classical phase space without using wavefunctions and operators.
In this regard, the projection property $W_{n\ell}\star W_{n\ell}=W_{n\ell}$ of Wigner
functions and the associativity and trace (or the so-called closedness) properties of
the star-product stand out. The latter reflects the fact that the integral of $f\star
g$ all over the phase space is equal to the integral of $fg$ (and therefore of $g\star
f$), where $f$ and $g$ are two arbitrary phase space functions.

\subsection{Inner Product, CBS Inequality and State Functional in a Phase Space}

In view of the above remarks we define a positive semidefinite
Hermitian inner product on the (cartesian  product of) linear space
of all phase space functions as follows
\begin{eqnarray}
<f|g>_{n\ell}=\frac{1}{h^{2}}\int_{\mathbb{R}^{4}}(\bar{f}\star g)W_{n\ell}dV\;.
\end{eqnarray}
This in particular implies the phase space analogue of the Cauchy-Bunyakowsky-Schwartz
(CBS) inequality
\begin{eqnarray}
 <f|f>_{n\ell}<g|g>_{n\ell}\geq |<f|g>_{n\ell}|^{2}\;.
\end{eqnarray}
 The basic mathematical structure and tool that lead to these two important
 relations are the associative $\star$-algebra structure of the phase
 space functions which admits the usual complex conjugation as an
 (anti)involution
\begin{eqnarray}
\overline{(\bar{f})}=f,\;\overline{(f\star g)}=\bar{g}\star \bar{f} \;,\nonumber
\end{eqnarray}
and the expectation (or mean) value function that can be defined as
\begin{eqnarray}
<f>_{n\ell}=\frac{1}{h^{2}}\int_{\mathbb{R}^{4}}f\star W_{n\ell}
dV=\frac{1}{h^{2}}\int_{\mathbb{R}^{4}}f W_{n\ell} dV\;.
\end{eqnarray}

In algebraic terms the expectation value is a state functional $s=s_{n\ell}$ on the
$\star$-algebra of the phase space functions which in our case simply reads as
$s(f)=<f>_{n\ell}$ . Defining properties of $s$ are that it is a complex linear
function and obey the relations \cite{Przanowski,Bordemann}
\begin{eqnarray}
s(1)=1\;,\quad s(\bar{f}\star f)\geq 0 \;,\nonumber
\end{eqnarray}
 The first relation is guarantied by the normalization of Wigner function
 and making use of the above mentioned properties the second can be verified
 as follows (see also \cite{Zachos4})
\begin{eqnarray}
\int_{\mathbb{R}^{4}}(\bar{f}\star f) W_{n\ell} dV&=&\int_{\mathbb{R}^{4}}(\bar{f}\star
f)\star
(W_{n\ell}\star W_{n\ell}) dV\nonumber\\
&=&\int_{\mathbb{R}^{4}}\bar{f}(f\star W_{n\ell}\star W_{n\ell})
dV\nonumber\\
&=&\int_{\mathbb{R}^{4}}(f\star W_{n\ell})\star(W_{n\ell}\star \bar{f}) dV=
\int_{\mathbb{R}^{4}}|f\star W_{n\ell}|^{2}dV\geq 0 \;.\nonumber
\end{eqnarray}
Note that $s(\bar{f}\star f)=0$ implies $f\star W_{n\ell}=0$ instead of $f=0$. That is
why the inner product
\begin{eqnarray}
s(\bar{f}\star g)=<f|g>_{n\ell} \;,\nonumber
\end{eqnarray}
is, in the case of fixed $ W_{n\ell}$, positive semidefinite.

It should be stressed that instead of Wigner functions any normalized and real-valued
projection function (or a set of such functions) that describes state space of a given
system can equally well be used in all these constructions. Two instances of this fact
will appear in the last two subsections.

\subsection{General Form of Uncertainty Relation For Two Functions}

Adapting the inequality (5.2) to the phase space functions
\begin{eqnarray}
\delta f=f-<f>\;,\quad \delta g=g-<g> \;,\nonumber
\end{eqnarray}
we have
\begin{eqnarray}
 (\Delta f)^{2} (\Delta g)^{2}\geq|<\delta f|\delta g>|^{2}\;,
\end{eqnarray}
where
\begin{eqnarray}
(\Delta f)^{2}=<\delta f|\delta f>=<f|f>-<f><\bar{f}> \;,\nonumber
\end{eqnarray}
is the variance of $f$ in a state described by the Wigner function $W$ whose quantum
numbers are, for simplicity, suppressed. Since $\{\delta f,\delta g \}_{M}=\{ f, g
\}_{M}$ we also have
\begin{eqnarray}
\delta f\star \delta g=\frac{1}{2}\{ f, g \}_{M}+\frac{1}{2}\{\delta f,\delta g
\}_{+M}\;,
\end{eqnarray}
where $\{ , \}_{+M}$ denotes the anti-Moyal bracket. Provided that
$f$ and $g$ are real-valued, at the right hand side of (5.5) the
first term is a pure imaginary-valued and the second is a
real-valued function. Analyzing the right hand side of (5.4) in view
of (5.5) we obtain
\begin{eqnarray}
 (\Delta f)^{2} (\Delta g)^{2}\geq -\frac{1}{4}< \{f,g\}_{M}>^{2}+
\frac{1}{4}< \{\delta f,\delta g\}_{+M}>^{2}
 \;.
\end{eqnarray}
This is the most general form of the phase space uncertainty relation for two
real-valued phase space functions which corresponds to the well-known
Robertson-Schr\"{o}dinger uncertainty relation.

\subsection{Uncertainty Products for the Phase Space Coordinates}

Recalling the fact that $(x_{\star})^{k}=x^{k}$ for the phase space
coordinates, the moments of coordinates can, by Eqs. (4.6), (4.17)
and (5.3), be directly computed from
\begin{eqnarray}
<x_{j}^{k}>_{n\ell}=
\frac{1}{h^{2}}\int_{-\infty}^{\infty}x_{j}^{k}P_{n\ell}(x_{j})dx_{j}\;.
\end{eqnarray}
This shows the importance of marginal probability densities in
explicit calculations. The fact that $P_{n\ell}(x_{i})$ are even
functions implies that the moments of coordinates are zero for odd
integer values of $k$. Using
\begin{eqnarray}
x^{2}=\frac{1}{4}[2H_{0}(x)+H_{2}(x)]\;,\nonumber
\end{eqnarray}
Eqs. (4.17) and the orthogonality relation (4.21) in Eq. (5.7) we
find
\begin{eqnarray}
<q_{1}^{2}>_{n\ell}&=&\frac{\gamma^{3}N_{q}}{2h^{2}}\pi^{1/2}\sum_{j=0}^{n}
\sum_{k=0}^{\ell}A_{n\ell
jk}(4\delta_{n+\ell-j-k,1}+\delta_{n+\ell-j-k,0}) \nonumber\\
&=&\frac{\gamma^{2}}{8}(4A_{n\;\ell\; n\; \ell-1}+4A_{n\;\ell\; n-1\;
\ell}+A_{n\;\ell\; n\; \ell})\;. \nonumber
\end{eqnarray}

From Eq. (4.16) we have
\begin{eqnarray}
A_{n\;\ell\; n\; \ell-1}=\ell\;,\quad A_{n\;\ell\; n-1\; \ell}=n\;,\quad A_{n\;\ell\;
n\; \ell}=4\;. \nonumber
\end{eqnarray}
and by substituting these into above relation we obtain
\begin{eqnarray}
<q_{1}^{2}>_{n\ell}=\frac{1}{2}\gamma^{2}(n+\ell+1)=<q_{2}^{2}>_{n\ell}\;. \nonumber
\end{eqnarray}
The results for momentum are found to be
\begin{eqnarray}
<p_{1}^{2}>_{n\ell}=\frac{1}{2}(\frac{\hbar}{\gamma})^{2}(n+\ell+1)=<p_{2}^{2}>_{n\ell}\;.
\nonumber
\end{eqnarray}
As the first moment of coordinates vanishes these are equal to the variances $(\Delta
x)_{n\ell}^{2}=<x^{2}>_{n\ell}-<x>^{2}_{n\ell}$. Therefore
\begin{eqnarray}
(\Delta q_{j})_{n\ell}(\Delta p_{j})_{n\ell}=\frac{1}{2}\hbar(n+\ell+1)\;,\quad
j=1,2\;.
\end{eqnarray}

These are the same as that can be found for Landau levels in the Schr\"{o}dinger
formulation. For all Landau levels the uncertainty products respect the lower bound
inequality $(\Delta q_{j})_{n\ell}(\Delta p_{j})_{n\ell}\geq \hbar/2$ and the equality
is satisfied only for the ground state. As we are about to see this is an expected
result since the ground state Wigner function is a coherent state corresponding to
$\alpha_{1}=0=\alpha_{2}$.

\subsection{Uncertainty Products for Phase Space Standard Coherent States}

The real and normalized coherent states of the Landau system defined by
\begin{eqnarray}
G_{s}=D_{\alpha_{1}\alpha_{2}}\star W_{0}\star \bar{D}_{\alpha_{1}\alpha_{2}}\;,
\end{eqnarray}
satisfy the same normalization and (by Eq. (3.8)) projection
properties of $W_{0}$
\begin{eqnarray}
\int_{R^{4}}G_{s}dV=\int_{R^{4}} W_{0}dV=h^{2}\;,\quad G_{s}\star
G_{s}=G_{s}\;.
\end{eqnarray}
We can therefore define the inner product and analyze the
uncertainty structures in a coherent state as well. In that case
expectation value will be defined as
\begin{eqnarray}
<f>_{cs}=\frac{1}{h^{2}}\int_{\mathbb{R}^{4}}f\star G_{s} dV\;,
\end{eqnarray}
where the subscripts $cs$ stand for coherent state.

By direct computation we obtain from (3.2)
\begin{eqnarray}
<a>_{cs}&=&\alpha_{1}\;,\; <b>_{cs}=\alpha_{2} \;,\nonumber\\
<\bar{a}>_{cs}&=&\bar{\alpha}_{1}\;,\;<\bar{b}>_{cs}=\bar{\alpha}_{2} \;,\nonumber
\end{eqnarray}
Noting, in view of Eqs. (2.1-2.2), that
\begin{eqnarray}
q_{1}&=& i\frac{\gamma}{2}[(a-b)-(\bar{a}-\bar{b})]\;,\nonumber\\
p_{1}&=& \frac{m\gamma \omega}{4}[(a-b)+(\bar{a}-\bar{b})]\;,\nonumber
\end{eqnarray}
we also get
\begin{eqnarray}
<q_{1}>_{cs}&=& -\gamma(\alpha_{1I}-\alpha_{2I})\;,\\
\ <p_{1}>_{cs}&=& \frac{m\gamma \omega}{2}(\alpha_{1R}-\alpha_{2R})\;,
\end{eqnarray}
where $\alpha_{kR}$ and  $\alpha_{kI}$ stand, respectively, for the
real and imaginary parts of $\alpha_{k}$. Using Eq. (3.2) and
$\{a-b,\bar{a}-\bar{b}\}=2$ we can write
\begin{eqnarray}
q^{2}_{1}\star G_{s}&=& -(\frac{\gamma}{2})^{2}[(a-b)-(\bar{a}-\bar{b})]\star
[(\alpha_{1}-\alpha_{2})-(\bar{a}-\bar{b})]\star G_{s}\;\nonumber\\
&=&-(\frac{\gamma}{2})^{2}[(\alpha_{1}-\alpha_{2})^{2}-2-2(\alpha_{1}-\alpha_{2})
(\bar{a}-\bar{b})+(\bar{a}-\bar{b})\star (\bar{a}-\bar{b})]\star G_{s}\;.\nonumber
\end{eqnarray}
Integrating  all over the phase space gives, by Eqs. (3.2), (5.10)
and trace property
\begin{eqnarray}
<q^{2}_{1}>_{cs}=\frac{\gamma^{2}}{2}+ \gamma^{2} (\alpha_{1I}-\alpha_{2I})^{2}\;.
\end{eqnarray}
By Eq. (5.12) this implies $(\Delta q_{1})^{2}_{cs}=\gamma^{2}/2$
and similar calculations gives $(\Delta
p_{1})^{2}_{cs}=\hbar^{2}/2\gamma^{2}$. These led us to $(\Delta
q_{1})_{cs}(\Delta p_{1})_{cs}=\hbar/2$. The same equality holds for
the other canonical pair.

As a result $G_{s}$ represents phase space coherent state with the minimum uncertainty
for all values of $\alpha_{1}$ and $\alpha_{2}$ and the variances of coordinates in
these states are equal to their values in the ground state.

\subsection{The Case of Generalized Coherent States}

Had we defined, in the sense of Perelomov, the generalized coherent states by applying
the displacement function to Wigner function $W_{n\ell}$ such that
\begin{eqnarray}
G_{g}=D_{\alpha_{1}\alpha_{2}}\star W_{n\ell}\star \bar{D}_{\alpha_{1}\alpha_{2}}\;,
\end{eqnarray}
the variances of coordinates in such a state would have been the same as that
calculated in the state $W_{n\ell}$. Finally in this section we will prove that this
claim is a special case of a more general fact.

In view of Eqs. (3.8) and (3.9) we can write
\begin{eqnarray}
f\star G_{g}&=& f\star D_{\alpha_{1}\alpha_{2}}\star W_{n\ell}\star
\bar{D}_{\alpha_{1}\alpha_{2}}\;\nonumber\\
&=& D_{\alpha_{1}\alpha_{2}}\star (\bar{D}_{\alpha_{1}\alpha_{2}}\star f\star
D_{\alpha_{1}\alpha_{2}})\star W_{n\ell}\star \bar{D}_{\alpha_{1}\alpha_{2}}\;\nonumber\\
&=& D_{\alpha_{1}\alpha_{2}}\star(f^{\prime}\star W_{n\ell})\star
\bar{D}_{\alpha_{1}\alpha_{2}}\;\nonumber
\end{eqnarray}
where $f$ is a smooth arbitrary phase space function and $f^{\prime}$ is its displaced
function (see Eqs. (3.9) and (3.10)). Similarly for any $k$th star power of $f$ we have
\begin{eqnarray}
(f_{\star})^{k}\star G_{g}=
D_{\alpha_{1}\alpha_{2}}\star((f^{\prime})_{\star})^{k}\star W_{n\ell})\star
\bar{D}_{\alpha_{1}\alpha_{2}}\;,
\end{eqnarray}
where $k$ is any positive integer. By integrating both sides of
(5.16) and by defining
\begin{eqnarray}
<(f_{\star})^{k}>_{g}=\frac{1}{h^{2}}\int_{\mathbb{R}^{4}}(f_{\star})^{k}G_{g}dV
\;,\nonumber
\end{eqnarray}
we immediately get
\begin{eqnarray}
<(f_{\star})^{k}>_{g}=<(f^{\prime}_{\star})^{k}>_{n\ell} \;.\nonumber
\end{eqnarray}

We conclude as follows. The expectation value of any integer star power
of a smooth phase space function in a generalized phase space coherent
state is equal to the expectation value of the same star power of the
corresponding displaced function in the state corresponding to the
Wigner function used in defining the generalized coherent state.

\begin{acknowledgments}
This work was supported in part by the Scientific and Technical Research Council of
Turkey (T\"{U}B\.{I}TAK).
\end{acknowledgments}


\end{document}